\documentclass[amsmath, amssymb, aps, superscriptaddress, twocolumn] {revtex4-1}

\usepackage{graphicx}
\usepackage{color}
\usepackage{dcolumn}
\usepackage{bm}
\usepackage{hyperref}
\usepackage[utf8]{inputenc}
\usepackage{amsmath}
\usepackage{graphicx}

\begin{document}

\title{Magnetoelastic standing waves induced in UO$_{2}$ by microsecond magnetic field pulses}

\author{Rico Sch{\"o}nemann} 
\email{rschoenemann@lanl.gov}
\affiliation{MPA-MAGLAB, Los Alamos National Laboratory, Los Alamos, NM 87545, USA.}

\author{George Rodriguez}
\affiliation{Q-6, Los Alamos National Laboratory, Los Alamos, NM 87545, USA.}

\author{Dwight Rickel}
\affiliation{MPA-MAGLAB, Los Alamos National Laboratory, Los Alamos, NM 87545, USA.}

\author{Fedor Balakirev}
\affiliation{MPA-MAGLAB, Los Alamos National Laboratory, Los Alamos, NM 87545, USA.}

\author{Ross D. McDonald}
 \affiliation{MPA-MAGLAB, Los Alamos National Laboratory, Los Alamos, NM 87545, USA.}

\author{Jordan Evans}
 \affiliation{MPA-MAGLAB, Los Alamos National Laboratory, Los Alamos, NM 87545, USA.}

\author{Boris Maiorov}
 \affiliation{MPA-MAGLAB, Los Alamos National Laboratory, Los Alamos, NM 87545, USA.}

\author{Charles Paillard}
\affiliation{Universit\'e Paris-Saclay, CNRS, UMR8580, Lab SPMS, Cent Supelec, 8-10 Rue Joliot Curie, F-91190 Gif Sur Yvette, France}
\affiliation{Physics Department and Institute for Nanoscience and Engineering, University of Arkansas, Fayetteville, Arkansas 72701, USA}
 
\author{Laurent Bellaiche}
\affiliation{Physics Department and Institute for Nanoscience and Engineering, University of Arkansas, Fayetteville, Arkansas 72701, USA}

\author{Myron B. Salamon}
\affiliation{MPA-MAGLAB, Los Alamos National Laboratory, Los Alamos, NM 87545, USA.}

\author{Krzysztof Gofryk}
\affiliation{Idaho National Laboratory, Idaho Falls, Idaho 83415, USA}

\author{Marcelo Jaime}
\email{mjaime@lanl.gov}
\affiliation{MPA-MAGLAB, Los Alamos National Laboratory, Los Alamos, NM 87545, USA.}
\affiliation{Physikalisch-Technische Bundesanstalt, 38116 Brauschweig, Germany}
\date{\today}

\begin{abstract}

Magnetoelastic measurements in the piezomagnetic antiferromagnet UO$_{2}$ were performed via the fiber Bragg grating method in magnetic fields up to $150\,\mathrm{T}$ generated by a single-turn coil setup. We show that in short timescales, order of a few micro seconds, pulsed-magnetic fields excite mechanical resonances at temperatures ranging from $10\,\mathrm{K}$ to $300\,\mathrm{K}$, in the paramagnetic as well as within the robust antiferromagnetic state of the material. These resonances, which are barely attenuated within the 100 ms observations, are attributed to the strong magnetoelastic coupling in UO$_{2}$ combined with the high crystallographic quality of the single crystal samples. They compare well with mechanical resonances obtained by a resonant ultrasound technique and superimpose on the known non-monotonic magnetostriction background. A clear phase-shift of $\pi$ in the lattice oscillations is, unexpectedly, observed in the antiferromagnetic state when the magnetic field overcomes the piezomagnetic switch-field $H_c \simeq -18\,\mathrm{T}$. We further present simulations and a theoretical argument to explain the observed phenomena.

\end{abstract}

\maketitle

\section{Introduction}

The antiferromagnetic (AFM) insulator Uranium dioxide UO$_{2}$ has been the subject of extensive research during the last decades predominantly due to its widespread use as nuclear fuel in pressurized heavy water reactors. Besides efforts to understand the unusually poor thermal conductivity of UO$_{2}$ which impacts its performance as nuclear fuel \cite{gofryk_anisotropic_2014}, a recent magnetostriction study in pulsed magnetic fields to $92\,\mathrm{T}$ uncovered linear magnetostriction in UO$_{2}$ \cite{jaime_piezomagnetism_2017} - a hallmark of piezomagnetism. 

Piezomagnetism is characterized by the induction of a magnetic polarization by application of mechanical strain, which, in the case of UO$_{2}$, is enabled by broken time-reversal symmetry in the 3-$k$ antiferromagnetic structure which emerges below $T_{\mathrm{N}} = 30.8\,\mathrm{K}$ \cite{burlet_neutron_1986, caciuffo_magnetic_1999, blackburn_spherical_2005, caciuffo_anisotropic_2007} and is accompanied by a Jahn-Teller distortion of the oxygen cage \cite{ giannozzi_theoretical_1987, ikushima_first-order_2001, santini_multipolar_2009, caciuffo_multipolar_2011}. This also leads to a complex hysteretic magnetoelastic memory behavior where magnetic domain switching occurs at fields around $\pm 18\,\mathrm{T}$ at $T = 2.5\,\mathrm{K}$. Interestingly, the very large applied magnetic fields proved unable to suppress the AFM state that sets in at $T_{\mathrm{N}}$ \cite{jaime_piezomagnetism_2017}. These earlier results provide direct evidence for the unusually high energy scale of spin-lattice interactions, and call for further studies in higher magnetic fields.

\begin{figure*}
\includegraphics{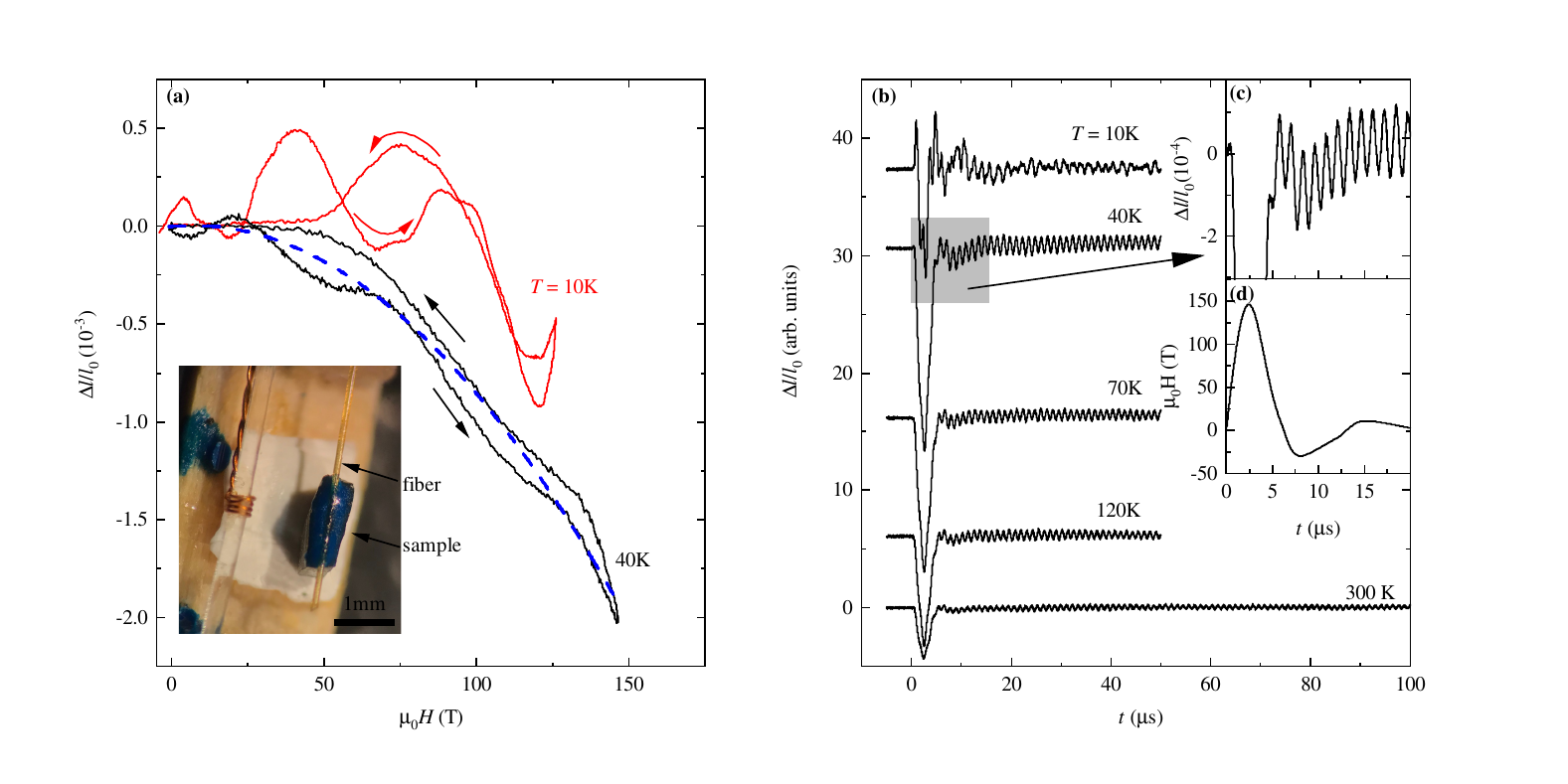}
\caption{(a) Uniaxial magnetostriction $\Delta l/l_{0}$ of a UO$_{2}$ single crystal along the [111] axis for $T = 10\,\mathrm{K}$ (red line) and $40\,\mathrm{K}$ (black line). The dashed blue line is a 2nd order polynomial fit of the $40\,\mathrm{K}$ data representing the expected magnetostriction behavior. The arrows marking up and down sweep of the magnetic field. The inset shows a picture of the UO$_{2}$ single crystal attached to the FBG-furnished fiber. Note that the sample is not physically attached to the sample holder, other than through the $125\,\mu m$ optical fiber. A copper coil, located next to the sample was used to measure the magnetic field. (b) $\Delta l/l_{0}$ of UO$_{2}$ as a function of time for pulses at different temperatures between 10 and $300\,\mathrm{K}$. The spike in $\Delta l/l_{0}$ during the first few microseconds is caused by the magnetostriction of the sample (shown in (a)) due to the changing magnetic field during the field pulse (d), whereas the oscillatory part of the signal continues after the field decayed to zero. A detailed view of the oscillations at $40\,\mathrm{K}$ is shown in the inset (c) as well as the magnetic field vs. time during the first $20\,\mathrm{\mu s}$ (d).}
\label{fig:figure1}
\end{figure*}

Here we present axial magnetostriction data obtained in a UO$_2$ single crystal in magnetic fields to $150\,\mathrm{T}$. These ultra-high fields were produced by single-turn coil pulsed resistive magnets and applied along the [111] crystallographic axis at various temperatures between $10\,\mathrm{K}$ and room temperature. We see, at all temperatures, a dominant negative magnetostriction proportional to the square of the applied field accompanied by unexpectedly strong oscillations that establish a mechanical resonance in the sample virtually instantly upon delivery of the ultra-fast, $10^{2}\,\mathrm{T/\mu \mathrm{s}}$, magnetic field rate-of-change. The oscillations observed quickly set, well within a single oscillation period, are long-lasting due to very low losses, with frequencies in the hundreds of kilohertz that match mechanical resonances obtained with a resonant ultrasound spectroscopy (RUS) technique \cite{balakirev_resonant_2019}. When the sample temperature is reduced, the frequencies soften, consistent with observations in studies of the UO$_2$ elastic constant $c_{44}$ as a function of temperature \cite{brandt_temperature_1967}.  When the magnetic field is applied at temperatures $T < 30.8\,\mathrm{K}$ in the AFM state, the magnetic field changes sign to a negative field magnitude (a characteristic of destructive magnets) in excess of the UO$_2$ AFM domain-switch-field of $\simeq$ -18 T. This negative field that follows a positive field pulse exposes yet another unexpected result, namely a $\pi$ (180$^o$) phase-shift in the magnetoelastic oscillations. We use a driven harmonic oscillator and a analytical model to shed light on the origin of our findings.

\section{Methods}

In this work, the magnetostriction signal of UO$_{2}$ was measured with a $100\,\mathrm{MHz}$ coherent pulse fiber Bragg interrogation method. The setup is driven by a modelocked pulsed $90\,\mathrm{fs}$ Er laser with a $100\,\mathrm{MHz}$ repetition rate and allows interrogation speed on the $10\,\mathrm{ns}$ scale. This method offers a faster readout rate than traditional fiber Bragg grating (FBG) interrogation systems which operate in the range of several kHz \cite{jaime_FBG_dilatometry_review_2017}. The UO$_{2}$ single crystal was attached to the optical fiber using an epoxy encapsulant with the crystallographic [111] axis aligned parallel to the fiber and the magnetic field. A picture of the sample is shown as an inset in figure \ref{fig:figure1}(a). Details about the FBG setup can be found in Ref. \cite{rodriguez_coherent_2015}. During the field pulse, the induced voltage in a small copper coil, located in close proximity to the sample (see picture in figure \ref{fig:figure1}(a)), was used to measure the magnetic field.

\begin{figure}
\includegraphics{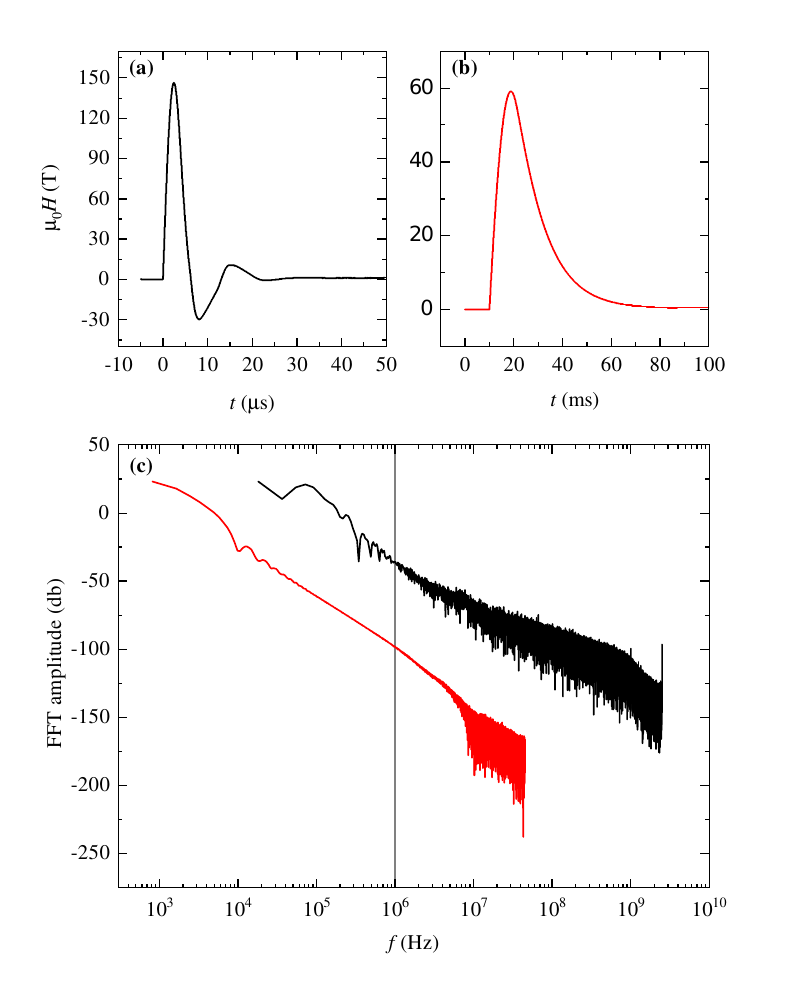}
\caption{Magnetic field vs. time of a $146\,\mathrm{T}$ pulse in a single-turn coil (a) and a $60\,\mathrm{T}$ pulse in a non-destructive short pulse magnet (b). (c) Fourier transformations of the field vs. time curves shown in (a, b). FFT's of the short pulse magnet/single-turn pulses are depicted by red/black lines respectively.}
\label{fig:figure2}
\end{figure}

The magnetic field was generated with a semi destructive capacitor-driven single-turn coil magnet system at the National High Magnetic Field Laboratory's pulse field facility at Los Alamos National Laboratory. The system is designed for fields up to $300\,\mathrm{T}$ with a rise time of approximately $2.5\,\mathrm{\mu s}$. Note that for magnetic fields in the region of $200\,\mathrm{T}$ and above, damage to the cryostat and sample becomes increasingly likely. Therefore magnetic fields were limited to $\approx 150\,\mathrm{T}$ in this study, with a peak rate of change in the order of $10^{2}\,\mathrm{T/\mu \mathrm{s}}$. Further details about the single-turn coil setup are presented in Refs. \cite{mielke_experimental_2010, mielke_single_2006}. Optical measurement techniques, like the FBG method used here, are in general advantageous in single-turn experiments when compared to, $e.g.$, electrical capacitance-based dilatometry measurements predominantly due to optical fibers being impervious to the large induced voltages generated inside even small metallic loops caused by the large $\mathrm{d}B/\mathrm{d}t$, as well as the associated electromagnetic noise.

The measurement of the natural mechanical resonances of elastic vibration where several normal modes of the sample are determined, is obtained with a set of piezoelectric transducers using a technique known as resonant ultrasound spectroscopy. Here, one transducer serves as source of the tunable sinusoidal wave of frequency $f$ and the other serves as detector at the synchronous frequency of the sample’s response. The electronics and room temperature apparatus was described in detail by Balakirev \textit{et al.} \cite{balakirev_resonant_2019}. In our case, the transducer had an Al$_{2}$O$_{3}$ hemisphere that allows precise and reproducible point contact on desired positions of the crystal \cite{evans_determining_2021}. As a frequency scan is performed, a resonance peak is observed at each of the normal modes.  We performed resonant mode measurements on UO$_{2}$ single crystals alone as well with the $125\,\mathrm{\mu m}$ optical fiber attached.

\begin{figure}
\includegraphics{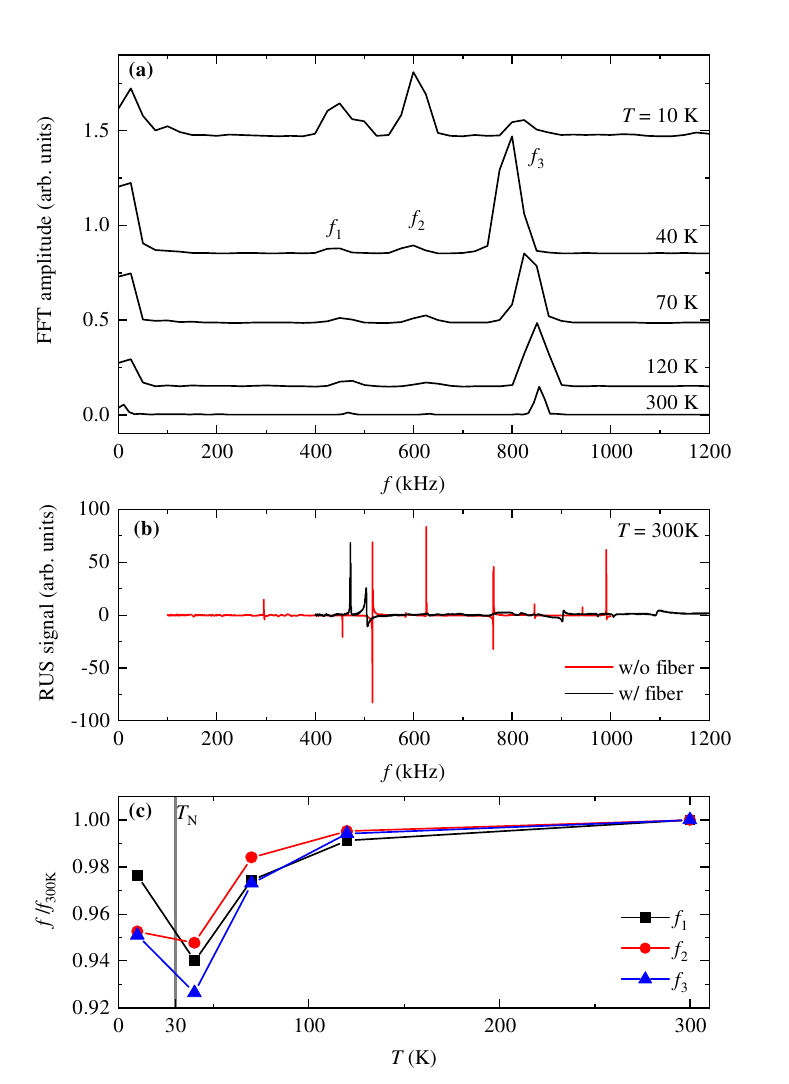}
\caption{(a) Fourier transforms of $\Delta l/l_{0}$ shown in figure \ref{fig:figure1}(b) displaying three dominant frequencies labeled as $f_{1}$, $f_{2}$ and $f_{3}$. The low frequency peaks originate from the background magnetostriction. (b) Resonant ultrasound spectroscopy (RUS) spectra at $T = 300\,\mathrm{K}$. The RUS spectra of the bare sample (black line) and the sample with the attached optical fiber (red line) were recorded along the [111] axis. (c) Normalized temperature dependence of $f_{1,2,3}$. }
\label{fig:figure3}
\end{figure}

\begin{figure*}
\includegraphics{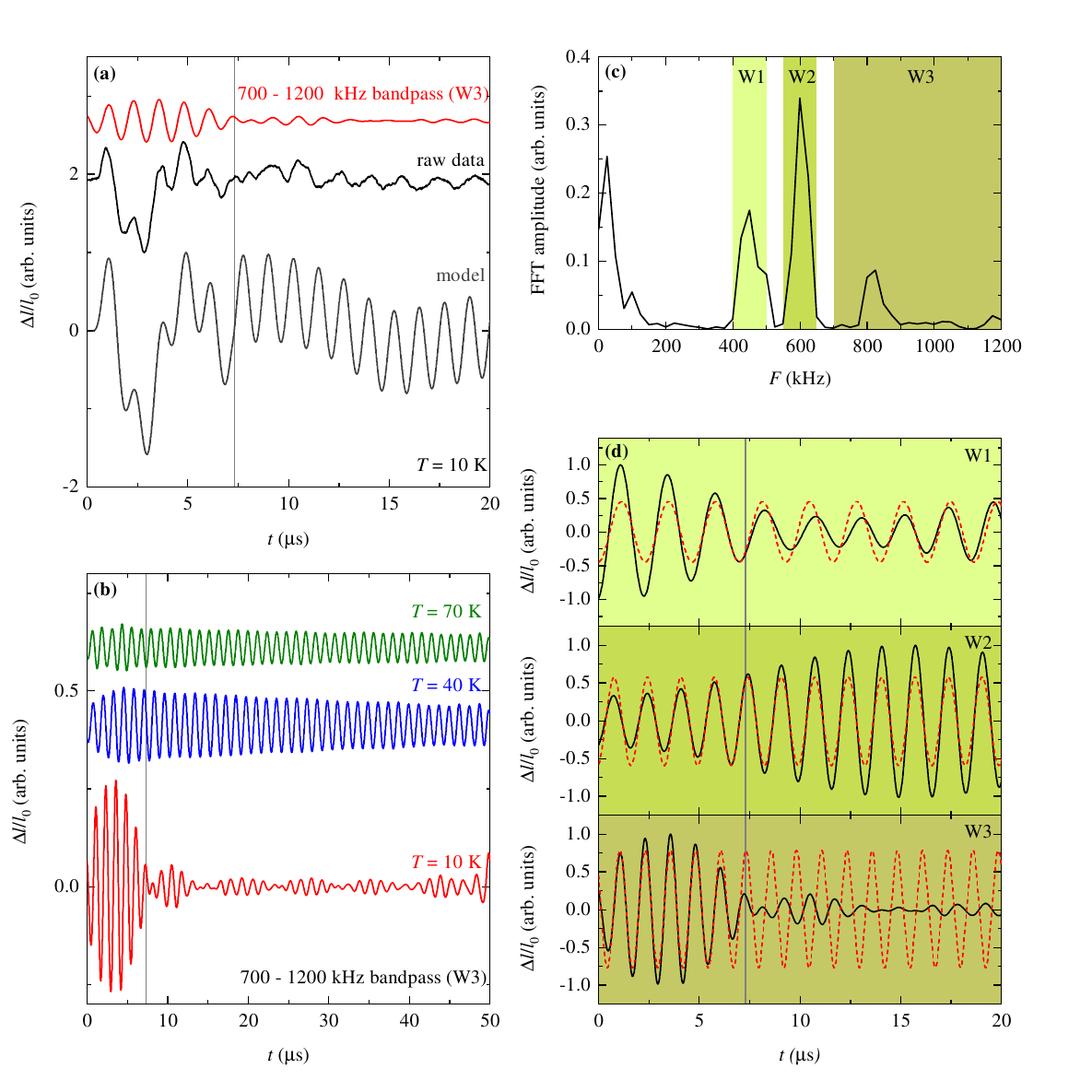}
\caption{(a) $\Delta l/l_{0}$ vs. time at $10\,\mathrm{K}$ with a peak field of $126\,\mathrm{T}$ before (black line) and after (red line) the 700-$1200\,\mathrm{kHz}$ band pass filter was applied. The experimental data is compared to a driven harmonic oscillator model with a characteristic frequency of $800\,\mathrm{kHz}$ matching the dominant oscillations seen during the field pulse in the magnetostriction signal (grey curve). The experimental data and the model show a $\pi$ phase shift around $7.3\,\mathrm{\mu s}$ marked by the grey vertical line. (b) Magnetoelastic oscillations at 10, 40 and $70\,\mathrm{K}$ after application of a 700-$1200\,\mathrm{kHz}$ band pass filter to the experimental data. Curves are shifted for clarity in figures (a, b). (c) Fast Fourier Transform of $\Delta l/l_{0}$ at $T=10\,\mathrm{K}$. The windows of the band pass filter (W1, W2, W3) are highlighted in different colors. The filtered curves (black lines) for each window are shown in (d) and are compared with sinusoidal functions indicated as red dashed lines.}
\label{fig:figure4}
\end{figure*}

\section{Results}

Magnetostriction $\Delta l/l_{0}$ vs. magnetic field curves at $T$ = $10\,\mathrm{K}$ and $40\,\mathrm{K}$ are displayed in Fig. \ref{fig:figure1}. We observe an overall negative magnetostriction signal at high fields as expected from previous results in pulsed magnetic fields to $92\,\mathrm{T}$ \cite{jaime_piezomagnetism_2017} with no indication of suppression of the robust AFM order. However, the signal is highly hysteretic due to the large mechanical resonances that are superimposed on the magnetostriction signal. The magnetostriction signal itself roughly follows a second order polynomial field dependence (blue dashed line).

The field-induced mechanical resonances become clearer when $\Delta l/l_{0}$ is plotted as a function of time (Fig. \ref{fig:figure1}(b)). Oscillations start with the onset of the field pulse and persist during the entire data acquisition period ($100\,\mathrm{\mu s}$ at $300\,\mathrm{K}$, $50\,\mathrm{\mu s}$ for lower temperatures). The high quality of our UO$_2$ crystal is probably a key factor behind the low attenuation of the mechanical resonances observed in the experiment. This is validated by a large quality factor $Q = f/w$ found by RUS and ranging from 2500 to 5000. Here, $f$ is the frequency and $w$ the width of the resonance (fitted by a Lorenzian). The onset of the mechanical resonances is approximately instantaneous, which indicates that they arise as a response to the magnetic field change and the strong magnetoelastic coupling. Hence, it does not appear that this mechanism is triggered by the shock wave generated by the disintegration of the single turn coil which would need a few microseconds to reach the sample. A similar experiment run with identical interrogation parameters and a bare FBG sensor, i.e. with no sample attached to the fiber, yielded no detectable mechanical resonances.  

In order to understand the origin of the observed mechanical resonances, we compare the typical field vs. time profile of a $146\,\mathrm{T}$ pulse performed in a single turn coil (Fig. \ref{fig:figure2}(a)) with a $60\,\mathrm{T}$ shot in a short-pulse magnet (Fig. \ref{fig:figure2}(b)). The field generated by the short pulse magnet has a total duration of about $100\,\mathrm{ms}$ with a rise time of $10\,\mathrm{ms}$. The single turn coil on the other hand has a pulse duration in the order of $25\,\mathrm{\mu s}$ and a rise time of $2.5\,\mathrm{\mu s}$ ($\times$4000 faster compared to the non-destructive magnet). Furthermore, the field switches sign several times, referred to as magnetic field recoil, displaying a significantly less attenuated behavior than the short-pulse magnet. The extremely short timescales in the single turn pulse results in a shift of the Fast Fourier Transform (FFT) of the field pulse towards higher frequencies up to several MHz (i.e., high frequencies are $50\,\mathrm{dB}$ more intense) which is displayed in Fig. \ref{fig:figure2}(c). Thus, if the system under study is magnetic - the field-pulse itself can, by virtue of the strong magnetoelastic coupling present in UO$_{2}$ \cite{dolling_crystal_1965, dolling_observation_1966, brandt_temperature_1967}, excite mechanical resonances in the range of several $100\,\mathrm{kHz}$.

The FFT of $\Delta l/l_{0} (t)$ in UO$_2$ reveals three distinct frequencies labeled $f_{1}$, $f_{2}$ and $f_{3}$ for temperatures between 40 and $300\,\mathrm{K}$, as shown in Fig. \ref{fig:figure3}(a). All modes display a softening as the temperature is lowered and a stiffening below $T_{\mathrm{N}}$ (Fig. \ref{fig:figure3}(c)), in agreement with previous measurements of the elastic constants of UO$_{2}$ which show a similar behavior \cite{brandt_temperature_1967}. Three independent elastic constants $c_{11}$, $c_{12}$ and $c_{44}$ exist for a lattice with cubic symmetry. The distinct temperature dependence of all three mechanical resonance frequencies $f_{1-3}$ indicates that we probe predominantly resonances that are associated with $c_{44}$ \cite{brandt_temperature_1967}.  The observed frequencies are in agreement with RUS data obtained at $300\,\mathrm{K}$, shown in Fig. \ref{fig:figure3}(b). The RUS measurements reveal a rich spectrum of sharp resonances in the frequency range up to $1\,\mathrm{MHz}$. When the crystal is attached to the fiber, a broadening of resonance peaks is observed, as well as a decrease in amplitude, indicating a larger damping, product of a larger system consisting of the crystal, glue and fiber. Nevertheless, the overall range of mechanical resonance frequencies observed in the magnetostriction measurements is comparable with RUS spectra. In the FBG measurements we observe three distinct frequencies. The additional mechanical resonance frequencies present in the RUS spectra are potential resonances that are either completely damped by the attached fiber and encapsulant or they only have a small compressive component parallel to the fiber since the FBG method is less sensitive to sheer strain. Note that the RUS measurements on the bare UO$_{2}$ crystal were performed on a $\approx 2\,\mathrm{mm}$ longer sample than the FBG and RUS measurements with the optical fiber attached. The change of the sample geometry also affects the mechanical resonance frequencies.

Upon cooling the sample below the antiferromagnetic transition we observe a substantial change in the mechanical resonances when compared to temperatures above $T_{\mathrm{N}}$: 

(i) In the data recorded at $10\,\mathrm{K}$, within the AFM ordered phase, the resonances appear to be significantly damped after $t\approx 7\mu\mathrm{s}$, particularly the FFT amplitude of $f_{3} = 800\,\mathrm{kHz}$ is suppressed when compared to $f_{1}$ and $f_{2}$, in contrast to the Fourier transforms of the data sets above $T_{\mathrm{N}}$ where $f_{3}$ is clearly the dominant resonance (Fig. \ref{fig:figure3}(a)). The attenuation effect becomes more apparent when band pass filters are applied to the experimental data, isolating the individual resonances and removing the magnetostriction background (Fig. \ref{fig:figure4}(a, b, d)). We show that, when compared to $f_{3}$, the amplitude of lower resonances $f_{2}$ and $f_{3}$ does not display a drastic change for $t > 7\,\mathrm{\mu s}$. The beating pattern observed in $f_{3}$ indicates the presence of two resonances very close to each other, which we cannot easily resolve in the FFT. These details are also impacted by the lower and upper limits chosen for the band pass filter.

(ii) A $\pi$ phase shift can be observed in the mechanical resonances at $t \approx 7\,\mathrm{\mu s}$ which is accompanied by the observed attenuation effect. One can clearly identify the $\pi$ phase shift after a 700-$1200\,\mathrm{kHz}$ high pass filter was applied to the experimental data (Fig. \ref{fig:figure4}(d)). Interestingly only $f_{3}$ shows the phase shift and both lower resonances $f_{1}$ and $f_{2}$ seem not to be affected and follow a single sinusoidal function as shown in Fig. \ref{fig:figure4}(d)). 

\begin{figure}
\includegraphics{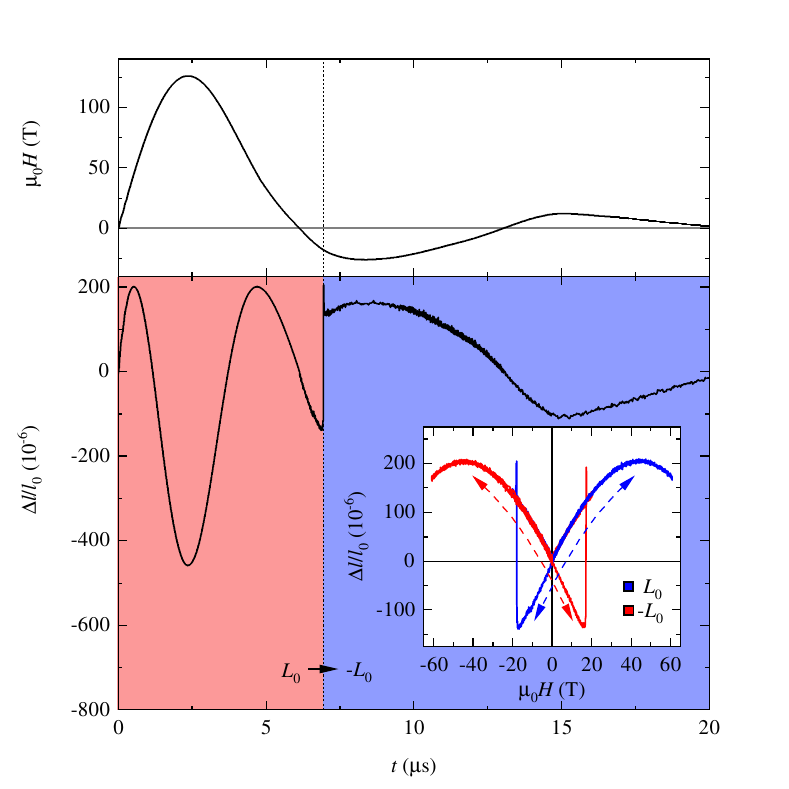}
\caption{Magnetic field (upper panel) and magnetostriction of UO$_{2}$ (lower panel) as a function of time. The magnetostriction data was mapped and extrapolated onto the single turn field profile using previously published pulse field data (Ref. \cite{jaime_piezomagnetism_2017}) shown in the inset. The vertical line at $t\approx7\,\mathrm{\mu s}$ separates the two states with positive (red) and negative (blue) AFM ordering vector $L_{0}$ by reaching the switching field of $\approx -18\,\mathrm{T}$ in this case.}
\label{fig:figure5}
\end{figure}

The phase shift around $t \approx 7\,\mathrm{\mu s}$ coincides with a magnetic field value of approximately $-18\,\mathrm{T}$ close to the field value where a abrupt sign change of the AFM ordering vector $L_{0}$ (as defined in Ref. \cite{Baryakhtar1985}) leads to a jump in the lattice distortion and the characteristic piezomagnetic butterfly that was reported in Ref. \cite{jaime_piezomagnetism_2017}. This effect is illustrated in Fig. \ref{fig:figure5} with the piezomagnetic butterfly shown in the inset. The origin of the phase shift can be found in the sudden reversal of  $L_{0}$ (red) to the $-L_{0}$ (blue) as we demonstrate in the following section.

\section{Hamiltonian}
Following Bar'yakhtar \textit{et al.}~\cite{Baryakhtar1985}, we denote as $\bm{L}_1$, $\bm{L}_2$ and $\bm{L}_3$ the different AFM vectors that describe the 3-$k$ order in UO\textsubscript{2}, and $\bm{M}$ the average magnetization. The magnetic unit cell below the AFM transition is made of 4 formula units, each formula unit carrying a magnetic moment $\bm{S}_1$, $\bm{S}_2$, $\bm{S}_3$, and $\bm{S}_4$. The above AFM vectors and magnetization can be expressed, in terms of the individual magnetic moments as:

\begin{eqnarray}
\bm{M} & = & \bm{S}_1+\bm{S}_2+\bm{S}_3+\bm{S}_4, \nonumber \\
\bm{L}_1 & = & \bm{S}_1+\bm{S}_2-\bm{S}_3-\bm{S}_4, \nonumber \\
\bm{L}_2 & = & \bm{S}_1-\bm{S}_2+\bm{S}_3-\bm{S}_4, \nonumber \\
\bm{L}_3 & = & \bm{S}_1-\bm{S}_2-\bm{S}_3+\bm{S}_4, \nonumber 
\end{eqnarray}

In the Landau approach, the thermodynamic potential must be able to describe both the paramagnetic and AFM phases. This is achieved by performing a polynomial expansion of the free energy whose terms respect the symmetry of the highest symmetry phase~\cite{Toledano}. Such terms have already been worked out in Ref.~\cite{Baryakhtar1985}. The Hamiltonian of the system can thus be written as:

\begin{equation}
\hat{H}  =  \int d^3\bm{r} \hat{\mathcal{H}}(\bm{r}),
\end{equation}

where $\hat{\mathcal{H}}$ can be found in the supplemental material. Since we are primarily interested in explaining the properties of the mechanical resonances and the $\pi$ phase shift in response to an applied pulsed magnetic fields, we make the simplifying following assumption: the magnetic response of the sample is primarily determined by the external magnetic field, and elastic vibrations do not affect it significantly. As a result, we write each magnetic quantity under the form $\bm{\mu} = \bm{\mu}_{eq} + \delta \bm{\mu}$ (with $\mu = L_1, L_2, L_3, M$). $\bm{\mu}_{eq}$ represents an equilibrium value, and $\delta \bm{\mu}$ the deviation away from that value, \textit{i.e.} the response to the magnetic pulse.

We then write the mechanical equations of motion, from which we retain  $\rho\ddot{u}_x$ 
	
	\begin{equation}
	\label{eq:ux}
		\rho \ddot{u}_x = \frac{\partial}{\partial x} \left( \frac{\partial \hat{\mathcal{H}}}{\partial \eta_1} \right) +  \frac{\partial}{\partial y} \left( \frac{\partial \hat{\mathcal{H}}}{\partial \eta_6} \right) +  \frac{\partial}{\partial z} \left( \frac{\partial \hat{\mathcal{H}}}{\partial \eta_5} \right),
	\end{equation}

	
	
	In the mechanical equations, $\rho$ is the volumic mass and $u_{\alpha}$ is the displacement field in direction $\alpha$. Since the derivations are lengthy, we will focus on the component $u_x$ of the displacement field which, after replacing with the Hamiltonian expression (see SI)
and assuming, as demonstrated in Ref. \cite{Baryakhtar1985}, that at equilibrium, $L_{1x}^{eq} = L_0$,  $L_{2y}^{eq} = L_0$ and $L_{3z}^{eq} = L_0$, we write $L_{1x} = L_0 + \delta L_{1x}$, $L_{1y} = \delta L_{1y}$, $L_{1z} = \delta L_{1z}$, etc. Similarly, Ref.~\cite{jaime_piezomagnetism_2017} shows that no magnetization seems to exist in the antiferromagnetic phase, so we write $M_x=\delta M_x$, etc. We also recall that $\eta_1 = \frac{\partial u_x}{\partial x}$, $\eta_5 = \frac{\partial u_x}{\partial z} +  \frac{\partial u_z}{\partial x}$,	
the equation can now be written, for linear terms, as


\begin{widetext}
\vspace{5mm}
\begin{eqnarray}
\rho \ddot{u}_x & \approx & c_{11} 
 \frac{\partial^2 u_x}{\partial x^2} + c_{12} \left( \frac{\partial^2 u_y}{\partial x \partial y} + \frac{\partial^2 u_z}{\partial x \partial z} \right) 
+ c_{44} \left( \frac{\partial^2 u_x}{\partial y^2} + \frac{\partial^2 u_y}{\partial y \partial x} + \frac{\partial^2 u_z}{\partial z \partial x} + \frac{\partial^2 u_x}{\partial z^2} \right) \nonumber \\ 
                          & + & L_0 (\lambda_1 + \lambda_1^{\prime}) \frac{\partial \delta L_{1x}}{\partial x} + \lambda_1^{\prime} L_0 \left( \frac{\partial \delta L_{2y}}{\partial x} + \frac{\partial \delta L_{3z}}{\partial x}  \right) 
                           +  \frac{\lambda}{2} L_0 \left( \frac{\partial \delta L_{1y}}{\partial y} + \frac{\partial \delta L_{2x}}{\partial y} + \frac{\partial \delta L_{1z}}{\partial z} + \frac{\partial \delta L_{3x}}{\partial z}  \right). \label{eq:ux_dev2}
\end{eqnarray}

\vspace{5mm}

In Equation~\ref{eq:ux_dev2}, the terms proportional to $\lambda_1$, $\lambda_1^{\prime}$ and $\lambda$ relate the linear change of shape of an antiferromagnetic uniform domain with respect to antiferromagnetic excitation.

If we perform a Fourier transform $u_x = \int \frac{d\bm{q}}{2\pi} \int \frac{d\omega}{2\pi} u_x(\bm{q},\omega) e^{i(\bm{q}.\bm{r} - \omega t)}$, etc., 
yielding:
\vspace{5mm}
\begin{eqnarray}
\left( c_{11} q_x^2 + c_{44} \left[ q_y^2 + q_z^2 \right] -\rho \omega^2 \right) u_x(\bm{q},\omega) & = & -(c_{44}+c_{12}) q_x \left[ q_y u_y(\bm{q},\omega) + q_z u_z(\bm{q},\omega) \right]  \nonumber \\
               & + & i q_x  {\color{red} L_0} \left[ (\lambda_1 + \lambda_1^{\prime}) \delta L_{1x}(\bm{q},\omega) + \lambda_1^{\prime}\left( \delta L_{2y}(\bm{q},\omega) + \delta L_{3z}(\bm{q},\omega) \right)  \right] \nonumber \\
               & + & \frac{\lambda}{2} {\color{red} L_0} \left( q_y\left[ \delta L_{1y}(\bm{q},\omega) + \delta L_{2x}(\bm{q},\omega)  \right] + q_z \left[ \delta L_{1z}(\bm{q},\omega) + \delta L_{3x}(\bm{q},\omega)  \right]  \right) \label{eq:ux_dev3}
\end{eqnarray}
\vspace{35mm}

\end{widetext}

We now have sets of coupled harmonic oscillators which are driven by a force which is proportional to $L_0$ (a cyclic permutation $x \rightarrow y \rightarrow z$ allows to get the equations for  the other components). In other words, given a proper change of basis, we can diagonalize this set of equations and write the displacement fields dynamical equations under the form 
\[ (\omega_0(\bm{q})^2 - \omega^2) u_1(\bm{q}, \omega) = F_1(\bm{q}, \omega), \]
with $F_1(\bm{q},\omega)$ being the force driving the oscillation of $u_1$ at pulsation $\omega$ with wavevector $\bm{q}$. This has an obvious solution, which is
\[ u_1(\bm{q}, \omega) = \frac{F_1(\bm{q}, \omega)}{\omega_0(\bm{q})^2 - \omega^2}.\]

We note that $F_1$ is proportional to $L_0$, the antiferromagnetic order parameter. It is then clear that upon reversal of the AFM order, $L_0 \longrightarrow -L_0$, the force applied on the set of harmonic oscillators reverses sign, \textit{i.e.} $F_1 \rightarrow -F_1 = e^{i\pi} F_1$ and thus a $\pi$ phase shift must be experienced in the elastic oscillatory response of the sample long enough after the pulse. Hence, sufficient switching of the AFM order is likely the cause of the phase shift observed in Fig. \ref{fig:figure4}. It is to be noted that the main energy couplings responsible for such an effect are quadratic in the AFM vectors and linear in strain; in other words, they are typical (antiferro)magnetostriction terms. We note that some of those energy couplings are the same ones from which piezomagnetism arises (see Equation 31-32 from Ref.~\cite{Baryakhtar1985}). 

Therefore, by assuming that the force driving the oscillations is proportional to the systems strain (shown in Fig. \ref{fig:figure5}) and fixing the frequency at $800\,\mathrm{kHz}$, we are able to model the experimental data with a simple driven harmonic oscillator. As depicted in Fig. \ref{fig:figure4}(a) this harmonic oscillator model reproduces the magnetostriction background as well as the $\pi$ phase shift in the $800\,\mathrm{kHz}$ oscillations. The attenuation observed in the oscillations after the switching of $L_{0}$ is not captured by the model and will be discussed in detail below. 

\section{Discussion}

A recent X-ray study on UO$_{2}$ single crystals evidences the presence of AFM domains and subsequently the coexistence of AFM phases $L_{0}$ and $-L_{0}$  connected by time-reversal even in magnetic fields beyond the piezomagnetic switching field \cite{antonio_piezomagnetic_2021}. This is also supported by magnetostriction measurements \cite{jaime_piezomagnetism_2017}, which show that the first pulse taken below $T_{\mathrm{N}}$ always has a smaller magnetostriction slope for fields below the switching field. This could be caused by the coexistence of all possible domains, with some contracting and some expanding as the field increases. In our measurements we observe a large attenuation effect around the switching field but oscillations seem not to be further damped afterwards. Therefore, the attenuation of $f_{3}$ appears to be caused by the coupling of the mechanical resonances to critical spin fluctuations and/or domain movement close to the piezomagnetic switching field which can lead to a significant attenuation of the mechanical resonances similar to the dramatically increased ultrasonic attenuation that was observed in UO$_{2}$ in the vicinity of the AFM phase transition \cite{brandt_temperature_1967}. The mechanical resonances $f_{1}$ and $f_{2}$ might have a predominantly transversal character which would explain the smaller amplitude and the absence of attenuation below $T_{N}$ since the longitudinal or compressive modes are expected to be more affected by spin fluctuations \cite{Yoshimasa1975}. For future experiments we plan to perform magnetocaloric measurements to detect possible heating effects at the switching field caused by dissipative processes like domain movement.

Another interesting point is that the $\pi$ phase shift only occurs in $f_{3}$. The effect is completely absent in $f_{2}$ and much less clear in $f_{1}$ which is only slightly out of phase when compared to the single sinusoidal function in the time interval between 0 and $20\,\mathrm{\mu s}$ (Fig. \ref{fig:figure4}(d)). As of now we do not have a conclusive argument on why the phase shift is only visible in $f_{3}$. Depending on the involved antiferromagnetic excitations and the anisotropy of the magneto-elastic couplings, longitudinal and transversal mechanical resonances can display different phase shifts. A possible way to test this in future experiments is to use a birefringent FBG which can yield an orthogonal biaxial strain response along two directions with polarization based probing techniques. 


We demonstrate that mechanical resonances can be a useful tool to detect otherwise-elusive AFM domain flips, and possibly also other types of crystallographic domain dynamics (e.g., in liquid crystals). On the other hand, our results indicate that mechanical resonances can also cause issues in experiments where they are unwanted. A mitigation strategy in experiments where excessive noise is prevalent could consist of clamping the sample as well as to conduct runs with different sample and/or sample-holders geometries and dimensions to minimize mechanical resonances triggered by the magnetic field. This phenomena is reminiscent of wire-motion resonances in electrical transport experiments performed in short-pulse magnets, which can be quite detrimental to the data quality and which effects are minimized by fixing the wires and in this way effectively shifting their resonances to frequencies outside of the experimental range of interest. 


\section{Summary}

We measured the lattice dilation along [111] for the first time up to 150T in UO$_2$, in the AFM as well as in the paramagnetic states. Surprisingly, the AFM state is robust against a field $\ge $ 120T at $10\,\mathrm{K}$, energy-wise $\sim4\times$ stronger than T$_N$ = 30K (if $g=2$). This result confirms the large energy scale for correlations in UO$_2$. We show that mechanical resonances can be induced virtually instantaneously via the magnetoelastic coupling in UO$_{2}$ by $\mu s$ field pulses generated with the single turn coil technique, making this material an interesting candidate for magneto-elastic transducers. We demonstrate the impact of the piezomagnetic switching in UO$_{2}$ on the standing wave indicated by a $\pi$ phase shift and a distinct mode dependent attenuation of the mechanical resonances. Our findings present a novel way to study magnetic dynamics in high magnetic fields and could have an impact on the interpretation of past and future data collected in experiments involving semi destructive pulsed magnetic fields as well practical implications, e.g., as a way to trigger resonators at faster speeds.

\section*{Acknowledgements}

A portion of this work was performed at the National High Magnetic Field Laboratory, which is supported by the National Science Foundation Cooperative Agreement No. DMR-1644779 and the state of Florida. R.S. acknowledges funding through the Seaborg Institute and the NHMFL UCGP program. C.P. and L.B thank the DARPA Grant No. HR0011-15-2-0038 (MATRIX program). K.G. acknowledges support from the US DOE BES Energy Frontier Research Centre "Thermal Energy Transport under Irradiation" (TETI). M.J. acknowledges support from the US DOE Basic Energy Science program through the project "Science at 100T" at LANL. MJ. and G.R. acknowledge support from the LANL Institute for Materials Science. 

\vspace{10mm}


\bibliography{UO2_SingleTurn}

\begin{thebibliography}{23}%
\makeatletter
\providecommand \@ifxundefined [1]{%
 \@ifx{#1\undefined}
}%
\providecommand \@ifnum [1]{%
 \ifnum #1\expandafter \@firstoftwo
 \else \expandafter \@secondoftwo
 \fi
}%
\providecommand \@ifx [1]{%
 \ifx #1\expandafter \@firstoftwo
 \else \expandafter \@secondoftwo
 \fi
}%
\providecommand \natexlab [1]{#1}%
\providecommand \enquote  [1]{``#1''}%
\providecommand \bibnamefont  [1]{#1}%
\providecommand \bibfnamefont [1]{#1}%
\providecommand \citenamefont [1]{#1}%
\providecommand \href@noop [0]{\@secondoftwo}%
\providecommand \href [0]{\begingroup \@sanitize@url \@href}%
\providecommand \@href[1]{\@@startlink{#1}\@@href}%
\providecommand \@@href[1]{\endgroup#1\@@endlink}%
\providecommand \@sanitize@url [0]{\catcode `\\12\catcode `\$12\catcode
  `\&12\catcode `\#12\catcode `\^12\catcode `\_12\catcode `\%12\relax}%
\providecommand \@@startlink[1]{}%
\providecommand \@@endlink[0]{}%
\providecommand \url  [0]{\begingroup\@sanitize@url \@url }%
\providecommand \@url [1]{\endgroup\@href {#1}{\urlprefix }}%
\providecommand \urlprefix  [0]{URL }%
\providecommand \Eprint [0]{\href }%
\providecommand \doibase [0]{http://dx.doi.org/}%
\providecommand \selectlanguage [0]{\@gobble}%
\providecommand \bibinfo  [0]{\@secondoftwo}%
\providecommand \bibfield  [0]{\@secondoftwo}%
\providecommand \translation [1]{[#1]}%
\providecommand \BibitemOpen [0]{}%
\providecommand \bibitemStop [0]{}%
\providecommand \bibitemNoStop [0]{.\EOS\space}%
\providecommand \EOS [0]{\spacefactor3000\relax}%
\providecommand \BibitemShut  [1]{\csname bibitem#1\endcsname}%
\let\auto@bib@innerbib\@empty
\bibitem [{\citenamefont {Gofryk}\ \emph {et~al.}(2014)\citenamefont {Gofryk},
  \citenamefont {Du}, \citenamefont {Stanek}, \citenamefont {Lashley},
  \citenamefont {Liu}, \citenamefont {Schulze}, \citenamefont {Smith},
  \citenamefont {Safarik}, \citenamefont {Byler}, \citenamefont {McClellan},
  \citenamefont {Uberuaga}, \citenamefont {Scott},\ and\ \citenamefont
  {Andersson}}]{gofryk_anisotropic_2014}%
  \BibitemOpen
  \bibfield  {author} {\bibinfo {author} {\bibfnamefont {K.}~\bibnamefont
  {Gofryk}}, \bibinfo {author} {\bibfnamefont {S.}~\bibnamefont {Du}}, \bibinfo
  {author} {\bibfnamefont {C.~R.}\ \bibnamefont {Stanek}}, \bibinfo {author}
  {\bibfnamefont {J.~C.}\ \bibnamefont {Lashley}}, \bibinfo {author}
  {\bibfnamefont {X.~Y.}\ \bibnamefont {Liu}}, \bibinfo {author} {\bibfnamefont
  {R.~K.}\ \bibnamefont {Schulze}}, \bibinfo {author} {\bibfnamefont {J.~L.}\
  \bibnamefont {Smith}}, \bibinfo {author} {\bibfnamefont {D.~J.}\ \bibnamefont
  {Safarik}}, \bibinfo {author} {\bibfnamefont {D.~D.}\ \bibnamefont {Byler}},
  \bibinfo {author} {\bibfnamefont {K.~J.}\ \bibnamefont {McClellan}}, \bibinfo
  {author} {\bibfnamefont {B.~P.}\ \bibnamefont {Uberuaga}}, \bibinfo {author}
  {\bibfnamefont {B.~L.}\ \bibnamefont {Scott}}, \ and\ \bibinfo {author}
  {\bibfnamefont {D.~A.}\ \bibnamefont {Andersson}},\ }\href {\doibase
  10.1038/ncomms5551} {\bibfield  {journal} {\bibinfo  {journal} {Nature
  Communications}\ }\textbf {\bibinfo {volume} {5}},\ \bibinfo {pages} {4551}
  (\bibinfo {year} {2014})}\BibitemShut {NoStop}%
\bibitem [{\citenamefont {Jaime}\ \emph
  {et~al.}(2017{\natexlab{a}})\citenamefont {Jaime}, \citenamefont {Saul},
  \citenamefont {Salamon}, \citenamefont {Zapf}, \citenamefont {Harrison},
  \citenamefont {Durakiewicz}, \citenamefont {Lashley}, \citenamefont
  {Andersson}, \citenamefont {Stanek}, \citenamefont {Smith},\ and\
  \citenamefont {Gofryk}}]{jaime_piezomagnetism_2017}%
  \BibitemOpen
  \bibfield  {author} {\bibinfo {author} {\bibfnamefont {M.}~\bibnamefont
  {Jaime}}, \bibinfo {author} {\bibfnamefont {A.}~\bibnamefont {Saul}},
  \bibinfo {author} {\bibfnamefont {M.}~\bibnamefont {Salamon}}, \bibinfo
  {author} {\bibfnamefont {V.~S.}\ \bibnamefont {Zapf}}, \bibinfo {author}
  {\bibfnamefont {N.}~\bibnamefont {Harrison}}, \bibinfo {author}
  {\bibfnamefont {T.}~\bibnamefont {Durakiewicz}}, \bibinfo {author}
  {\bibfnamefont {J.~C.}\ \bibnamefont {Lashley}}, \bibinfo {author}
  {\bibfnamefont {D.~A.}\ \bibnamefont {Andersson}}, \bibinfo {author}
  {\bibfnamefont {C.~R.}\ \bibnamefont {Stanek}}, \bibinfo {author}
  {\bibfnamefont {J.~L.}\ \bibnamefont {Smith}}, \ and\ \bibinfo {author}
  {\bibfnamefont {K.}~\bibnamefont {Gofryk}},\ }\href {\doibase
  10.1038/s41467-017-00096-4} {\bibfield  {journal} {\bibinfo  {journal}
  {Nature Communications}\ }\textbf {\bibinfo {volume} {8}},\ \bibinfo {pages}
  {99} (\bibinfo {year} {2017}{\natexlab{a}})}\BibitemShut {NoStop}%
\bibitem [{\citenamefont {Burlet}\ \emph {et~al.}(1986)\citenamefont {Burlet},
  \citenamefont {Rossat-Mignod}, \citenamefont {vuevel}, \citenamefont {Vogt},
  \citenamefont {Spirlet},\ and\ \citenamefont
  {Rebivant}}]{burlet_neutron_1986}%
  \BibitemOpen
  \bibfield  {author} {\bibinfo {author} {\bibfnamefont {P.}~\bibnamefont
  {Burlet}}, \bibinfo {author} {\bibfnamefont {J.}~\bibnamefont
  {Rossat-Mignod}}, \bibinfo {author} {\bibfnamefont {S.}~\bibnamefont
  {vuevel}}, \bibinfo {author} {\bibfnamefont {O.}~\bibnamefont {Vogt}},
  \bibinfo {author} {\bibfnamefont {J.}~\bibnamefont {Spirlet}}, \ and\
  \bibinfo {author} {\bibfnamefont {J.}~\bibnamefont {Rebivant}},\ }\href
  {\doibase 10.1016/0022-5088(86)90521-7} {\bibfield  {journal} {\bibinfo
  {journal} {Journal of the Less Common Metals}\ }\textbf {\bibinfo {volume}
  {121}},\ \bibinfo {pages} {121} (\bibinfo {year} {1986})}\BibitemShut
  {NoStop}%
\bibitem [{\citenamefont {Caciuffo}\ \emph {et~al.}(1999)\citenamefont
  {Caciuffo}, \citenamefont {Amoretti}, \citenamefont {Santini}, \citenamefont
  {Lander}, \citenamefont {Kulda},\ and\ \citenamefont
  {Du~Plessis}}]{caciuffo_magnetic_1999}%
  \BibitemOpen
  \bibfield  {author} {\bibinfo {author} {\bibfnamefont {R.}~\bibnamefont
  {Caciuffo}}, \bibinfo {author} {\bibfnamefont {G.}~\bibnamefont {Amoretti}},
  \bibinfo {author} {\bibfnamefont {P.}~\bibnamefont {Santini}}, \bibinfo
  {author} {\bibfnamefont {G.~H.}\ \bibnamefont {Lander}}, \bibinfo {author}
  {\bibfnamefont {J.}~\bibnamefont {Kulda}}, \ and\ \bibinfo {author}
  {\bibfnamefont {P.~d.~V.}\ \bibnamefont {Du~Plessis}},\ }\href {\doibase
  10.1103/PhysRevB.59.13892} {\bibfield  {journal} {\bibinfo  {journal}
  {Physical Review B}\ }\textbf {\bibinfo {volume} {59}},\ \bibinfo {pages}
  {13892} (\bibinfo {year} {1999})}\BibitemShut {NoStop}%
\bibitem [{\citenamefont {Blackburn}\ \emph {et~al.}(2005)\citenamefont
  {Blackburn}, \citenamefont {Caciuffo}, \citenamefont {Magnani}, \citenamefont
  {Santini}, \citenamefont {Brown}, \citenamefont {Enderle},\ and\
  \citenamefont {Lander}}]{blackburn_spherical_2005}%
  \BibitemOpen
  \bibfield  {author} {\bibinfo {author} {\bibfnamefont {E.}~\bibnamefont
  {Blackburn}}, \bibinfo {author} {\bibfnamefont {R.}~\bibnamefont {Caciuffo}},
  \bibinfo {author} {\bibfnamefont {N.}~\bibnamefont {Magnani}}, \bibinfo
  {author} {\bibfnamefont {P.}~\bibnamefont {Santini}}, \bibinfo {author}
  {\bibfnamefont {P.~J.}\ \bibnamefont {Brown}}, \bibinfo {author}
  {\bibfnamefont {M.}~\bibnamefont {Enderle}}, \ and\ \bibinfo {author}
  {\bibfnamefont {G.~H.}\ \bibnamefont {Lander}},\ }\href {\doibase
  10.1103/PhysRevB.72.184411} {\bibfield  {journal} {\bibinfo  {journal}
  {Physical Review B}\ }\textbf {\bibinfo {volume} {72}},\ \bibinfo {pages}
  {184411} (\bibinfo {year} {2005})}\BibitemShut {NoStop}%
\bibitem [{\citenamefont {Caciuffo}\ \emph {et~al.}(2007)\citenamefont
  {Caciuffo}, \citenamefont {Magnani}, \citenamefont {Santini}, \citenamefont
  {Carretta}, \citenamefont {Amoretti}, \citenamefont {Blackburn},
  \citenamefont {Enderle}, \citenamefont {Brown},\ and\ \citenamefont
  {Lander}}]{caciuffo_anisotropic_2007}%
  \BibitemOpen
  \bibfield  {author} {\bibinfo {author} {\bibfnamefont {R.}~\bibnamefont
  {Caciuffo}}, \bibinfo {author} {\bibfnamefont {N.}~\bibnamefont {Magnani}},
  \bibinfo {author} {\bibfnamefont {P.}~\bibnamefont {Santini}}, \bibinfo
  {author} {\bibfnamefont {S.}~\bibnamefont {Carretta}}, \bibinfo {author}
  {\bibfnamefont {G.}~\bibnamefont {Amoretti}}, \bibinfo {author}
  {\bibfnamefont {E.}~\bibnamefont {Blackburn}}, \bibinfo {author}
  {\bibfnamefont {M.}~\bibnamefont {Enderle}}, \bibinfo {author} {\bibfnamefont
  {P.}~\bibnamefont {Brown}}, \ and\ \bibinfo {author} {\bibfnamefont
  {G.}~\bibnamefont {Lander}},\ }\href {\doibase 10.1016/j.jmmm.2006.10.536}
  {\bibfield  {journal} {\bibinfo  {journal} {Journal of Magnetism and Magnetic
  Materials}\ }\textbf {\bibinfo {volume} {310}},\ \bibinfo {pages} {1698}
  (\bibinfo {year} {2007})}\BibitemShut {NoStop}%
\bibitem [{\citenamefont {Giannozzi}\ and\ \citenamefont
  {Erdös}(1987)}]{giannozzi_theoretical_1987}%
  \BibitemOpen
  \bibfield  {author} {\bibinfo {author} {\bibfnamefont {P.}~\bibnamefont
  {Giannozzi}}\ and\ \bibinfo {author} {\bibfnamefont {P.}~\bibnamefont
  {Erdös}},\ }\href {\doibase 10.1016/0304-8853(87)90722-0} {\bibfield
  {journal} {\bibinfo  {journal} {Journal of Magnetism and Magnetic Materials}\
  }\textbf {\bibinfo {volume} {67}},\ \bibinfo {pages} {75} (\bibinfo {year}
  {1987})}\BibitemShut {NoStop}%
\bibitem [{\citenamefont {Ikushima}\ \emph {et~al.}(2001)\citenamefont
  {Ikushima}, \citenamefont {Tsutsui}, \citenamefont {Haga}, \citenamefont
  {Yasuoka}, \citenamefont {Walstedt}, \citenamefont {Masaki}, \citenamefont
  {Nakamura}, \citenamefont {Nasu},\ and\ \citenamefont
  {Ōnuki}}]{ikushima_first-order_2001}%
  \BibitemOpen
  \bibfield  {author} {\bibinfo {author} {\bibfnamefont {K.}~\bibnamefont
  {Ikushima}}, \bibinfo {author} {\bibfnamefont {S.}~\bibnamefont {Tsutsui}},
  \bibinfo {author} {\bibfnamefont {Y.}~\bibnamefont {Haga}}, \bibinfo {author}
  {\bibfnamefont {H.}~\bibnamefont {Yasuoka}}, \bibinfo {author} {\bibfnamefont
  {R.~E.}\ \bibnamefont {Walstedt}}, \bibinfo {author} {\bibfnamefont {N.~M.}\
  \bibnamefont {Masaki}}, \bibinfo {author} {\bibfnamefont {A.}~\bibnamefont
  {Nakamura}}, \bibinfo {author} {\bibfnamefont {S.}~\bibnamefont {Nasu}}, \
  and\ \bibinfo {author} {\bibfnamefont {Y.}~\bibnamefont {Ōnuki}},\ }\href
  {\doibase 10.1103/PhysRevB.63.104404} {\bibfield  {journal} {\bibinfo
  {journal} {Physical Review B}\ }\textbf {\bibinfo {volume} {63}},\ \bibinfo
  {pages} {104404} (\bibinfo {year} {2001})}\BibitemShut {NoStop}%
\bibitem [{\citenamefont {Santini}\ \emph {et~al.}(2009)\citenamefont
  {Santini}, \citenamefont {Carretta}, \citenamefont {Amoretti}, \citenamefont
  {Caciuffo}, \citenamefont {Magnani},\ and\ \citenamefont
  {Lander}}]{santini_multipolar_2009}%
  \BibitemOpen
  \bibfield  {author} {\bibinfo {author} {\bibfnamefont {P.}~\bibnamefont
  {Santini}}, \bibinfo {author} {\bibfnamefont {S.}~\bibnamefont {Carretta}},
  \bibinfo {author} {\bibfnamefont {G.}~\bibnamefont {Amoretti}}, \bibinfo
  {author} {\bibfnamefont {R.}~\bibnamefont {Caciuffo}}, \bibinfo {author}
  {\bibfnamefont {N.}~\bibnamefont {Magnani}}, \ and\ \bibinfo {author}
  {\bibfnamefont {G.~H.}\ \bibnamefont {Lander}},\ }\href {\doibase
  10.1103/RevModPhys.81.807} {\bibfield  {journal} {\bibinfo  {journal}
  {Reviews of Modern Physics}\ }\textbf {\bibinfo {volume} {81}},\ \bibinfo
  {pages} {807} (\bibinfo {year} {2009})}\BibitemShut {NoStop}%
\bibitem [{\citenamefont {Caciuffo}\ \emph {et~al.}(2011)\citenamefont
  {Caciuffo}, \citenamefont {Santini}, \citenamefont {Carretta}, \citenamefont
  {Amoretti}, \citenamefont {Hiess}, \citenamefont {Magnani}, \citenamefont
  {Regnault},\ and\ \citenamefont {Lander}}]{caciuffo_multipolar_2011}%
  \BibitemOpen
  \bibfield  {author} {\bibinfo {author} {\bibfnamefont {R.}~\bibnamefont
  {Caciuffo}}, \bibinfo {author} {\bibfnamefont {P.}~\bibnamefont {Santini}},
  \bibinfo {author} {\bibfnamefont {S.}~\bibnamefont {Carretta}}, \bibinfo
  {author} {\bibfnamefont {G.}~\bibnamefont {Amoretti}}, \bibinfo {author}
  {\bibfnamefont {A.}~\bibnamefont {Hiess}}, \bibinfo {author} {\bibfnamefont
  {N.}~\bibnamefont {Magnani}}, \bibinfo {author} {\bibfnamefont {L.-P.}\
  \bibnamefont {Regnault}}, \ and\ \bibinfo {author} {\bibfnamefont {G.~H.}\
  \bibnamefont {Lander}},\ }\href {\doibase 10.1103/PhysRevB.84.104409}
  {\bibfield  {journal} {\bibinfo  {journal} {Physical Review B}\ }\textbf
  {\bibinfo {volume} {84}},\ \bibinfo {pages} {104409} (\bibinfo {year}
  {2011})}\BibitemShut {NoStop}%
\bibitem [{\citenamefont {Balakirev}\ \emph {et~al.}(2019)\citenamefont
  {Balakirev}, \citenamefont {Ennaceur}, \citenamefont {Migliori},
  \citenamefont {Maiorov},\ and\ \citenamefont
  {Migliori}}]{balakirev_resonant_2019}%
  \BibitemOpen
  \bibfield  {author} {\bibinfo {author} {\bibfnamefont {F.~F.}\ \bibnamefont
  {Balakirev}}, \bibinfo {author} {\bibfnamefont {S.~M.}\ \bibnamefont
  {Ennaceur}}, \bibinfo {author} {\bibfnamefont {R.~J.}\ \bibnamefont
  {Migliori}}, \bibinfo {author} {\bibfnamefont {B.}~\bibnamefont {Maiorov}}, \
  and\ \bibinfo {author} {\bibfnamefont {A.}~\bibnamefont {Migliori}},\ }\href
  {\doibase 10.1063/1.5123165} {\bibfield  {journal} {\bibinfo  {journal}
  {Review of Scientific Instruments}\ }\textbf {\bibinfo {volume} {90}},\
  \bibinfo {pages} {121401} (\bibinfo {year} {2019})}\BibitemShut {NoStop}%
\bibitem [{\citenamefont {Brandt}\ and\ \citenamefont
  {Walker}(1967)}]{brandt_temperature_1967}%
  \BibitemOpen
  \bibfield  {author} {\bibinfo {author} {\bibfnamefont {O.~G.}\ \bibnamefont
  {Brandt}}\ and\ \bibinfo {author} {\bibfnamefont {C.~T.}\ \bibnamefont
  {Walker}},\ }\href {\doibase 10.1103/PhysRevLett.18.11} {\bibfield  {journal}
  {\bibinfo  {journal} {Physical Review Letters}\ }\textbf {\bibinfo {volume}
  {18}},\ \bibinfo {pages} {11} (\bibinfo {year} {1967})}\BibitemShut {NoStop}%
\bibitem [{\citenamefont {Jaime}\ \emph
  {et~al.}(2017{\natexlab{b}})\citenamefont {Jaime}, \citenamefont
  {Corvalan~Moya}, \citenamefont {Weickert}, \citenamefont {Zapf},
  \citenamefont {Balakirev}, \citenamefont {Wartenbe}, \citenamefont {Rosa},
  \citenamefont {Betts}, \citenamefont {Rodriguez}, \citenamefont {Crooker},\
  and\ \citenamefont {Daou}}]{jaime_FBG_dilatometry_review_2017}%
  \BibitemOpen
  \bibfield  {author} {\bibinfo {author} {\bibfnamefont {M.}~\bibnamefont
  {Jaime}}, \bibinfo {author} {\bibfnamefont {C.}~\bibnamefont
  {Corvalan~Moya}}, \bibinfo {author} {\bibfnamefont {F.}~\bibnamefont
  {Weickert}}, \bibinfo {author} {\bibfnamefont {V.}~\bibnamefont {Zapf}},
  \bibinfo {author} {\bibfnamefont {F.~F.}\ \bibnamefont {Balakirev}}, \bibinfo
  {author} {\bibfnamefont {M.}~\bibnamefont {Wartenbe}}, \bibinfo {author}
  {\bibfnamefont {P.~F.~S.}\ \bibnamefont {Rosa}}, \bibinfo {author}
  {\bibfnamefont {J.~B.}\ \bibnamefont {Betts}}, \bibinfo {author}
  {\bibfnamefont {G.}~\bibnamefont {Rodriguez}}, \bibinfo {author}
  {\bibfnamefont {S.~A.}\ \bibnamefont {Crooker}}, \ and\ \bibinfo {author}
  {\bibfnamefont {R.}~\bibnamefont {Daou}},\ }\href {\doibase
  10.3390/s17112572} {\bibfield  {journal} {\bibinfo  {journal} {Sensors}\
  }\textbf {\bibinfo {volume} {17}},\ \bibinfo {pages} {2572} (\bibinfo {year}
  {2017}{\natexlab{b}})}\BibitemShut {NoStop}%
\bibitem [{\citenamefont {Rodriguez}\ \emph {et~al.}(2015)\citenamefont
  {Rodriguez}, \citenamefont {Jaime}, \citenamefont {Balakirev}, \citenamefont
  {Mielke}, \citenamefont {Azad}, \citenamefont {Marshall}, \citenamefont
  {La~Lone}, \citenamefont {Henson},\ and\ \citenamefont
  {Smilowitz}}]{rodriguez_coherent_2015}%
  \BibitemOpen
  \bibfield  {author} {\bibinfo {author} {\bibfnamefont {G.}~\bibnamefont
  {Rodriguez}}, \bibinfo {author} {\bibfnamefont {M.}~\bibnamefont {Jaime}},
  \bibinfo {author} {\bibfnamefont {F.}~\bibnamefont {Balakirev}}, \bibinfo
  {author} {\bibfnamefont {C.~H.}\ \bibnamefont {Mielke}}, \bibinfo {author}
  {\bibfnamefont {A.}~\bibnamefont {Azad}}, \bibinfo {author} {\bibfnamefont
  {B.}~\bibnamefont {Marshall}}, \bibinfo {author} {\bibfnamefont {B.~M.}\
  \bibnamefont {La~Lone}}, \bibinfo {author} {\bibfnamefont {B.}~\bibnamefont
  {Henson}}, \ and\ \bibinfo {author} {\bibfnamefont {L.}~\bibnamefont
  {Smilowitz}},\ }\href {\doibase 10.1364/OE.23.014219} {\bibfield  {journal}
  {\bibinfo  {journal} {Optics Express}\ }\textbf {\bibinfo {volume} {23}},\
  \bibinfo {pages} {14219} (\bibinfo {year} {2015})}\BibitemShut {NoStop}%
\bibitem [{\citenamefont {Mielke}\ and\ \citenamefont
  {Novac}(2010)}]{mielke_experimental_2010}%
  \BibitemOpen
  \bibfield  {author} {\bibinfo {author} {\bibfnamefont {C.~H.}\ \bibnamefont
  {Mielke}}\ and\ \bibinfo {author} {\bibfnamefont {B.~M.}\ \bibnamefont
  {Novac}},\ }\href {\doibase 10.1109/TPS.2010.2049507} {\bibfield  {journal}
  {\bibinfo  {journal} {IEEE Transactions on Plasma Science}\ }\textbf
  {\bibinfo {volume} {38}},\ \bibinfo {pages} {1739} (\bibinfo {year}
  {2010})}\BibitemShut {NoStop}%
\bibitem [{\citenamefont {Mielke}\ and\ \citenamefont
  {McDonald}(2006)}]{mielke_single_2006}%
  \BibitemOpen
  \bibfield  {author} {\bibinfo {author} {\bibfnamefont {C.~H.}\ \bibnamefont
  {Mielke}}\ and\ \bibinfo {author} {\bibfnamefont {R.~D.}\ \bibnamefont
  {McDonald}},\ }in\ \href {\doibase 10.1109/MEGAGUSS.2006.4530682} {\emph
  {\bibinfo {booktitle} {2006 {IEEE} {International} {Conference} on
  {Magagauss} {Magnetic} {Field} {Generation} and {Related} {Topics}}}}\
  (\bibinfo  {publisher} {IEEE},\ \bibinfo {address} {Santa Fe, NM, USA},\
  \bibinfo {year} {2006})\ pp.\ \bibinfo {pages} {227--231}\BibitemShut
  {NoStop}%
\bibitem [{\citenamefont {Evans}\ \emph {et~al.}(2021)\citenamefont {Evans},
  \citenamefont {Sturtevant}, \citenamefont {Clausen}, \citenamefont {Vogel},
  \citenamefont {Balakirev}, \citenamefont {Betts}, \citenamefont {Capolungo},
  \citenamefont {Lebensohn},\ and\ \citenamefont
  {Maiorov}}]{evans_determining_2021}%
  \BibitemOpen
  \bibfield  {author} {\bibinfo {author} {\bibfnamefont {J.~A.}\ \bibnamefont
  {Evans}}, \bibinfo {author} {\bibfnamefont {B.~T.}\ \bibnamefont
  {Sturtevant}}, \bibinfo {author} {\bibfnamefont {B.}~\bibnamefont {Clausen}},
  \bibinfo {author} {\bibfnamefont {S.~C.}\ \bibnamefont {Vogel}}, \bibinfo
  {author} {\bibfnamefont {F.~F.}\ \bibnamefont {Balakirev}}, \bibinfo {author}
  {\bibfnamefont {J.~B.}\ \bibnamefont {Betts}}, \bibinfo {author}
  {\bibfnamefont {L.}~\bibnamefont {Capolungo}}, \bibinfo {author}
  {\bibfnamefont {R.~A.}\ \bibnamefont {Lebensohn}}, \ and\ \bibinfo {author}
  {\bibfnamefont {B.}~\bibnamefont {Maiorov}},\ }\href {\doibase
  10.1007/s10853-021-05827-z} {\bibfield  {journal} {\bibinfo  {journal}
  {Journal of Materials Science}\ }\textbf {\bibinfo {volume} {56}},\ \bibinfo
  {pages} {10053} (\bibinfo {year} {2021})}\BibitemShut {NoStop}%
\bibitem [{\citenamefont {Dolling}\ \emph {et~al.}(1965)\citenamefont
  {Dolling}, \citenamefont {Cowley},\ and\ \citenamefont
  {Woods}}]{dolling_crystal_1965}%
  \BibitemOpen
  \bibfield  {author} {\bibinfo {author} {\bibfnamefont {G.}~\bibnamefont
  {Dolling}}, \bibinfo {author} {\bibfnamefont {R.~A.}\ \bibnamefont {Cowley}},
  \ and\ \bibinfo {author} {\bibfnamefont {A.~D.~B.}\ \bibnamefont {Woods}},\
  }\href {\doibase 10.1139/p65-135} {\bibfield  {journal} {\bibinfo  {journal}
  {Canadian Journal of Physics}\ }\textbf {\bibinfo {volume} {43}},\ \bibinfo
  {pages} {1397} (\bibinfo {year} {1965})}\BibitemShut {NoStop}%
\bibitem [{\citenamefont {Dolling}\ and\ \citenamefont
  {Cowley}(1966)}]{dolling_observation_1966}%
  \BibitemOpen
  \bibfield  {author} {\bibinfo {author} {\bibfnamefont {G.}~\bibnamefont
  {Dolling}}\ and\ \bibinfo {author} {\bibfnamefont {R.~A.}\ \bibnamefont
  {Cowley}},\ }\href {\doibase 10.1103/PhysRevLett.16.683} {\bibfield
  {journal} {\bibinfo  {journal} {Physical Review Letters}\ }\textbf {\bibinfo
  {volume} {16}},\ \bibinfo {pages} {683} (\bibinfo {year} {1966})}\BibitemShut
  {NoStop}%
\bibitem [{\citenamefont {Bar'yakhtar}\ \emph {et~al.}(1985)\citenamefont
  {Bar'yakhtar}, \citenamefont {Vitebskii},\ and\ \citenamefont
  {Yablonskii}}]{Baryakhtar1985}%
  \BibitemOpen
  \bibfield  {author} {\bibinfo {author} {\bibfnamefont {V.~G.}\ \bibnamefont
  {Bar'yakhtar}}, \bibinfo {author} {\bibfnamefont {I.~M.}\ \bibnamefont
  {Vitebskii}}, \ and\ \bibinfo {author} {\bibfnamefont {D.~A.}\ \bibnamefont
  {Yablonskii}},\ }\href
  {http://www.jetp.ac.ru/cgi-bin/e/index/e/62/1/p108?a=list} {\bibfield
  {journal} {\bibinfo  {journal} {Sov. Phys. JETP}\ }\textbf {\bibinfo {volume}
  {62}},\ \bibinfo {pages} {108} (\bibinfo {year} {1985})}\BibitemShut
  {NoStop}%
\bibitem [{\citenamefont {Tol\'{e}dano}\ and\ \citenamefont
  {Tol\'{e}dano}(1987)}]{Toledano}%
  \BibitemOpen
  \bibfield  {author} {\bibinfo {author} {\bibfnamefont {J.~C.}\ \bibnamefont
  {Tol\'{e}dano}}\ and\ \bibinfo {author} {\bibfnamefont {P.}~\bibnamefont
  {Tol\'{e}dano}},\ }\href@noop {} {\emph {\bibinfo {title} {The Landau Theory
  of Phase Transitions}}}\ (\bibinfo {year} {1987})\BibitemShut {NoStop}%
\bibitem [{\citenamefont {Antonio}\ \emph {et~al.}(2021)\citenamefont
  {Antonio}, \citenamefont {Weiss}, \citenamefont {Shanks}, \citenamefont
  {Ruff}, \citenamefont {Jaime}, \citenamefont {Saul}, \citenamefont
  {Swinburne}, \citenamefont {Salamon}, \citenamefont {Shrestha}, \citenamefont
  {Lavina}, \citenamefont {Koury}, \citenamefont {Gruner}, \citenamefont
  {Andersson}, \citenamefont {Stanek}, \citenamefont {Durakiewicz},
  \citenamefont {Smith}, \citenamefont {Islam},\ and\ \citenamefont
  {Gofryk}}]{antonio_piezomagnetic_2021}%
  \BibitemOpen
  \bibfield  {author} {\bibinfo {author} {\bibfnamefont {D.~J.}\ \bibnamefont
  {Antonio}}, \bibinfo {author} {\bibfnamefont {J.~T.}\ \bibnamefont {Weiss}},
  \bibinfo {author} {\bibfnamefont {K.~S.}\ \bibnamefont {Shanks}}, \bibinfo
  {author} {\bibfnamefont {J.~P.~C.}\ \bibnamefont {Ruff}}, \bibinfo {author}
  {\bibfnamefont {M.}~\bibnamefont {Jaime}}, \bibinfo {author} {\bibfnamefont
  {A.}~\bibnamefont {Saul}}, \bibinfo {author} {\bibfnamefont {T.}~\bibnamefont
  {Swinburne}}, \bibinfo {author} {\bibfnamefont {M.}~\bibnamefont {Salamon}},
  \bibinfo {author} {\bibfnamefont {K.}~\bibnamefont {Shrestha}}, \bibinfo
  {author} {\bibfnamefont {B.}~\bibnamefont {Lavina}}, \bibinfo {author}
  {\bibfnamefont {D.}~\bibnamefont {Koury}}, \bibinfo {author} {\bibfnamefont
  {S.~M.}\ \bibnamefont {Gruner}}, \bibinfo {author} {\bibfnamefont {D.~A.}\
  \bibnamefont {Andersson}}, \bibinfo {author} {\bibfnamefont {C.~R.}\
  \bibnamefont {Stanek}}, \bibinfo {author} {\bibfnamefont {T.}~\bibnamefont
  {Durakiewicz}}, \bibinfo {author} {\bibfnamefont {J.~L.}\ \bibnamefont
  {Smith}}, \bibinfo {author} {\bibfnamefont {Z.}~\bibnamefont {Islam}}, \ and\
  \bibinfo {author} {\bibfnamefont {K.}~\bibnamefont {Gofryk}},\ }\href
  {\doibase 10.1038/s43246-021-00121-6} {\bibfield  {journal} {\bibinfo
  {journal} {Communications Materials}\ }\textbf {\bibinfo {volume} {2}},\
  \bibinfo {pages} {17} (\bibinfo {year} {2021})}\BibitemShut {NoStop}%
\bibitem [{\citenamefont {Itoh}(1975)}]{Yoshimasa1975}%
  \BibitemOpen
  \bibfield  {author} {\bibinfo {author} {\bibfnamefont {Y.}~\bibnamefont
  {Itoh}},\ }\href {\doibase 10.1143/JPSJ.38.336} {\bibfield  {journal}
  {\bibinfo  {journal} {Journal of the Physical Society of Japan}\ }\textbf
  {\bibinfo {volume} {38}},\ \bibinfo {pages} {336} (\bibinfo {year}
  {1975})}\BibitemShut {NoStop}%
\end{thebibliography}%

\end{document}


\title{Supplemental Material}

\maketitle

\section{Hamiltonian}

As mentioned in the main text, the terms of the Hamiltonian can be derived from the polynomial expansion of the free energy and were worked out by Bar’yakhtar \textit{et al.}. They can can be written as:

\begin{eqnarray}
\hat{\mathcal{H}} & = & \frac{1}{2}\alpha_L^0 (T-T_N) \left[ L_{1x}^2 + L_{2y}^2 + L_{3z}^2 \right] + \frac{1}{4} \beta_L \left[ \bm{L}_1^4 + \bm{L}_2^4 + \bm{L}_3^4 \right] \nonumber \\
&  & +\frac{1}{2} \alpha_M M^2 + \frac{1}{2} \lambda_{ML} \left[ \left( \bm{M}.\bm{L}_1 \right)^2 + \left( \bm{M}.\bm{L}_2 \right)^2 + \left( \bm{M}.\bm{L}_3 \right)^2 \right] \nonumber \\
&  & + \frac{c_{11}}{2} \left[ \eta_1^2 + \eta_2^2 + \eta_3^2 \right] + c_{12} \left[ \eta_1 \eta_2 + \eta_2 \eta_3 + \eta_3 \eta_1 \right] + \frac{c_{44}}{2} \left[ \eta_4^2 + \eta_5^2 + \eta_6^2 \right] \nonumber \\
&  & + \frac{\lambda_1}{2} \left( L_{1x}^2 \eta_1 + L_{2y}^2 \eta_2 + L_{3z}^2 \eta_3 \right) + \frac{\lambda_1^{\prime}}{2} \left( L_{1x}^2 + L_{2y}^2 + L_{3z}^2 \right) \left( \eta_1 + \eta_2 + \eta_3 \right) \nonumber \\
&  & + \frac{\lambda_2}{2} \left( L_{1y} L_{1z} \eta_4 + L_{2x} L_{2z} \eta_5 + L_{3x} L_{3y} \eta_6 \right) + \frac{\lambda_3}{2} \left( M_x^2 \eta_1 + M_y^2 \eta_2 + M_z^2 \eta_3 \right) \nonumber \\
&  & + \frac{\lambda_3^{\prime}}{2} \left( M_y M_z \eta_4 + M_x M_z \eta_5 + M_x M_y \eta_6 \right) \nonumber \\
&  & + \frac{\lambda}{2} \left( \left[ L_{1x} L_{1y} + L_{2x} L_{2y} \right] \eta_6 + \left[ L_{1x} L_{1z} + L_{3x} L_{3z} \right] \eta_5 + \left[ L_{2y} L_{2z} + L_{3y} L_{3z} \right] \eta_4 \right) \nonumber \\
&  & + \frac{A}{2} \left[ 
\left( \frac{\partial \bm{L_{1}}}{\partial x} \right)^2 
+ \left( \frac{\partial \bm{L_{1}}}{\partial y} \right)^2 
+ \left( \frac{\partial \bm{L_{1}}}{\partial z} \right)^2 
+ \left( \frac{\partial \bm{L_{2}}}{\partial x} \right)^2 
+ \left( \frac{\partial \bm{L_{2}}}{\partial y} \right)^2 + \left( \frac{\partial \bm{L_{2}}}{\partial z} \right)^2 \right. \nonumber \\
& & + \left. \left( \frac{\partial \bm{L_{3}}}{\partial x} \right)^2 + \left( \frac{\partial \bm{L_{3}}}{\partial y} \right)^2 + \left( \frac{\partial \bm{L_{3}}}{\partial z} \right)^2 
\right] \nonumber \\
&  & + \frac{B}{2} \left[ \left( \frac{\partial \bm{L_{1}}}{\partial x} \right)^2 + \left( \frac{\partial \bm{L_{2}}}{\partial y} \right)^2 + \left( \frac{\partial \bm{L_{3}}}{\partial z} \right)^2 \right] \nonumber \\
&   & - \bm{M}.\bm{H}. \label{eq:Hamiltonian}
\end{eqnarray}

Here, $x, y$ and $z$ are oriented along the axes of the cubic cell of UO\textsubscript{2}, and $\eta_i$, $i=1,6$ are the corresponding strain using Voigt notation.